\documentclass{article}
\usepackage{bigsky2007}
\usepackage{graphicx}
\frompage{000} \topage{000}                                              
\def\rightmark{First Excited State in $^{23}$O}
\def\leftmark{N. Frank et al.}

\title{Observation of the First Excited State in $^{23}$O}
\authors{
{N. Frank$^{1,2,3}$, A. Schiller$^1$, T. Baumann$^1$, D. Bazin$^1$, J. Brown$^4$, \\ P. A. DeYoung$^5$, J. E. Finck$^6$, A. Gade$^1$, J. Hinnefeld$^7$, R. Howes$^8$, J.-L. Lecouey$^{1,a}$, B. Luther$^3$, W. A. Peters$^{1,2}$, H. Scheit$^1$, and \\ M. Thoennessen$^{1,2}$ %
}\\[2.812mm]
{\normalsize
\hspace*{-8pt}$^1$ National Superconducting Cyclotron Laboratory, \\ Michigan State
University, East Lansing, MI 48824\\[0.2ex]
\hspace*{-8pt}$^2$ Department of Physics \& Astronomy, \\ Michigan State University, East Lansing, MI 48824\\[0.2ex]
\hspace*{-8pt}$^3$ Department of Physics, Concordia College, Moorhead, MN 56562\\[0.2ex]
\hspace*{-8pt}$^4$ Department of Physics, Wabash College, Crawfordsville, IN 47933\\[0.2ex]
\hspace*{-8pt}$^5$ Department of Physics, Hope College, Holland, MI 49423\\[0.2ex]
\hspace*{-8pt}$^6$ Department of Physics, Central Michigan University, Mt.\ Pleasant, MI 48859\\[0.2ex]
\hspace*{-8pt}$^7$ Department of Physics \& Astronomy, \\ Indiana University at South Bend, South Bend, IN 46634\\[0.2ex]
\hspace*{-8pt}$^8$ Department of Physics, Marquette University, Milwaukee, WI 53201\\[0.2ex]
}}

\abstract{The first excited state in neutron-rich $^{23}$O was observed in a (2p1n) knock-out reaction from $^{26}$Ne
on a beryllium target at a beam energy of 86 MeV/$A$. The state is unbound with respect to neutron emission and was
reconstructed from the invariant mass from the $^{22}$O fragment and the neutron. It is unbound by 45(2)~keV corresponding to
an excitation energy of 2.8(1) MeV. The non-observation of further resonances implies a predominantly direct reaction
mechanism of the employed three-nucleon-removal reaction which suggests the assignment of the observed resonance to be the
$5/2^+$ hole state. }

\keyword{Invariant mass method, neutron-rich nuclei, direct reaction, knockout reactions }
\PACS{21.10.Pc, 23.90.+w, 25.60.Gc, 29.30.Hs }

\begin{document}

\maketitle
\setcounter{page}{1}

\section{Introduction}\label{intro}

It has been established that the traditional magic numbers can disappear for nuclei far from stability\cite{Nav00,Dom06}. At the same
time rearrangement of single particle orbits opens new gaps leading to new magic numbers \cite{Bro01}. In
order to search for and study these changes, several observables have to be measured, for example, binding energies and
the level structure of excited states. For very neutron-rich nuclei close to the neutron dripline none or only a few
bound excited states exist, and for nuclei beyond the dripline even the ground state is unbound with respect to neutron
emission. Thus, $\gamma$-ray spectroscopy is not feasible and other detection methods have to be utilized. In very
light nuclei these states can be populated with multiple-particle transfer reactions from stable beams
\cite{Kal00,Boh07}. Another method is $\beta$-delayed neutron spectroscopy where neutron-unbound excited states are
populated via $\beta$-decay from neutron-rich isotopes produced in fast fragmentation reactions \cite{Sum06,Sum07}.
However, some nuclei can only be studied with the method of neutron-decay spectroscopy, where the states of interest
are populated either directly by fast fragmentation reactions \cite{Dea87,Kry93,Tho99}, or by few nucleon knockout
reaction of secondary beams (which first were produced by fast fragmentation) \cite{Zin97,Che01}. The excited states
are then reconstructed from the invariant mass of the emitted neutron and coincident fragment.

\begin{figure}[b!]
\begin{center}
\includegraphics[totalheight=4.7cm]{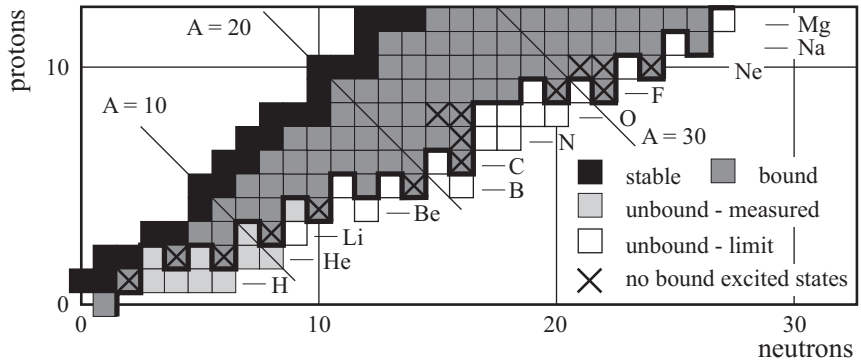}
\end{center}
\caption{Section of the chart of nuclei. Stable and neutron-bound isotopes are shown as black and dark grey squares,
respectively. Unbound nuclei beyond the dripline where some spectroscopy information is available and which have merely
been determined to be unbound are shown as light grey and white squares, respectively. Nuclei with no bound excited states are indicated by crosses.} \label{fig:chart}
\end{figure}

Figure \ref{fig:chart} shows the region of light, neutron-rich isotopes of the nuclear chart. Stable and neutron-bound
isotopes are shown as black and dark grey squares, respectively. Unbound nuclei beyond the dripline where some
spectroscopy information is available and which have merely been determined to be unbound are shown as light grey and
white squares, respectively. Nuclei with no bound excited states are indicated by crosses. Of these states only
$^{23}$O and $^{31}$Ne are accessible with $\beta$-delayed neutron spectroscopy.

In neutron-rich oxygen isotopes, the magic number $N = 20$ has vanished and new gaps at $N = 14$ and $N = 16$ have appeared
\cite{Oza00,Thi00}. No bound excited states have been observed in $^{23}$O or $^{24}$O \cite{Sta04} which is the last
bound oxygen isotope \cite{Lan85,Gui90,Tar97}. $^{23}$O is especially interesting because it is located between the new
shell gaps. While single particle states probe the $N = 16$ shell gap, single hole states probe the $N = 14$ gap. The
present study searches for the first excited state of $^{23}$O using the method of neutron decay spectroscopy and the
selective (2p1n) knockout reaction from $^{26}$Ne.

\section{Experimental Set-up}\label{setup}

The experiment was performed at the National Superconducting Cyclotron Laboratory at Michigan State University.
A 105~pnA primary beam of 140~MeV/$A$ $^{40}$Ar impinged on a 893~mg/cm$^2$ Be production target. The resulting
cocktail beam was purified with respect to the desired $^{26}$Ne at 86~MeV/$A$ using an achromatic 750-mg/cm$^2$-thick
acrylic wedge degrader in the A1900 fragment separator \cite{Mor03}. A purity of up to 93.2\% was achieved with a
$^{26}$Ne beam intensity of about 7000~pps. The contaminants (mainly $^{27}$Na and $^{29}$Mg) were separated
event-by-event in the off-line analysis by their different time-of-flight (ToF) from the extended focal plane to a
scintillator in front of the 721-mg/cm$^2$-thick Be reaction target. Positions and angles of incoming beam particles
were measured by two $15\times 15$~cm$^2$ position-sensitive parallel-plate avalanche counters (PPACs) with
$\sim$8~pads/cm, i.e., ${\mathrm{FWHM}}\approx 1.3$~mm. The presence of a quadrupole triplet downstream of the PPACs
translates this into a position resolution of impinging $^{26}$Ne particles on the target within a FWHM radius
of 2.4~mm.

Charged particles behind the target were bent $~43^\circ$ by the large-gap 4~Tm Sweeper Magnet \cite{Bir05}.
Two $30\times 30$~cm$^2$ cathode-readout drift chambers (CRDCs) provided position in the dispersive ($\sim$4~pads/cm,
${\mathrm{FWHM}}\approx 2.5$~mm) and non-dispersive (drift time, ${\mathrm{FWHM}}\approx 3.1$~mm) direction. The 1.87~m
distance between the CRDCs translates this into ${\mathrm{FWHM}}=2.4$~mrad angle resolution. Energy loss was determined
in a 65-cm-long ion chamber (IC) and a $40\times 40$~cm$^2$, 4.5-mm-thin plastic scintillator whose pulse-height signal
was corrected for position. Energy loss was used to separate reaction products with different $Z$. The thin
scintillator also gave ToF of reaction products from the target. This, together with the total kinetic energy measurement in a 15-cm-thick plastic scintillator, provided isotopic separation. Further details can be found in
\cite{Fra06,Sch07}.

Beam-velocity neutrons were detected in the Modular Neutron Array (MoNA) \cite{Lut03,Bau05} at a distance of 8.2~m from
the target with an intrinsic efficiency of $\sim 70\%$. MoNA consists of $9\times 16$ stacked 2-m-long plastic
scintillator bars which are read out on both ends by photomultiplier tubes (PMT). The bars are mounted horizontally and
perpendicular to the beam axis. Position along the vertical and along the beam axis is determined within the thickness
of one bar (10~cm). Horizontal position and neutron ToF are determined by the time difference and the mean time,
respectively, of the two PMT signals which yield resolutions of ${\mathrm{FWHM}}\approx 12$~cm and
${\mathrm{FWHM}}\approx 0.24$~ns, see also \cite{Sch07,Pet07}.

The decay energies of resonances are reconstructed by the invariant mass method. To do this, the relativistic four-momentum
vectors of the neutron and fragment are reconstructed at the point of breakup. For neutrons, position and ToF
resolution translate into angle and energy resolution of ${\mathrm{FWHM}}=19$~mrad and ${\mathrm{FWHM}}=3.8$~MeV,
respectively. The angle and energy of fragments in coincidence with neutrons were reconstructed behind the 
target based on the ion-optical properties of the Sweeper Magnet using a novel method which takes into account the
position at the target in the dispersive direction \cite{Fra07}.
The reconstructed position in the non-dispersive direction serves thereby as a check on an event-by-event basis against
the same information obtained from forward tracking from the PPACs through the quadrupole triplet. The angle and energy
resolution of the fragments behind the target are ${\mathrm{FWHM}}=6.4$~mrad and
${\mathrm{FWHM}}=0.9$~MeV/$A$, respectively. The average energy loss of the fragment through half of the 
target is added to approximate the relativistic four-momentum vector of the fragment at the average breakup point.

\section{Results}\label{results}
\begin{figure}[tb!]
\begin{center}
\includegraphics[totalheight=5.5cm]{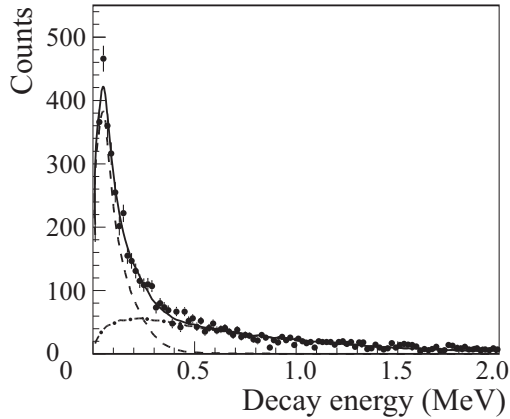}
\end{center}
\caption{Decay-energy spectra of $^{23}$O$^*$ (data points). The thick solid curve corresponds to
the sum of the dashed (simulated resonant contribution) and dash-dotted
(simulated thermal model) curves and agrees well with the data.}
\label{fig:decay}
\end{figure}

Figure \ref{fig:decay} shows the decay energy spectrum for $^{23}$O reconstructed from neutrons in coincidence with
$^{22}$O fragments. The data were compared with simulations based on a Glauber reaction model including angle
straggling of the fragments in the target as well as detector resolutions. The data in the figure are
best described by a Breit-Wigner resonance contribution (dashed curve) with $E_r=45(2)$~keV and an energy-independent
single-particle width of $\Gamma\sim 0.1$~keV plus a contribution from a beam-velocity source of Maxwellian-distributed neutrons (a
thermal model) with $T\sim 0.7$~MeV (dash-dotted curve). The decay-energy resolution is dominated by the neutron-angle
resolution and contributions due to the target thickness. The thermal model has been included in the fit
as a phenomenological description for non-resonant contribution to the spectrum.

The decay energy ($45(2)$~keV) of the resonance can be related to an excitation energy ($2.79(13)$~MeV) by adding the
neutron-separation energy $S_n=2.74(13)$~MeV of $^{23}$O \cite{Aud03} under the assumption that the decay populates the
ground state of $^{22}$O. The other alternative, that this decay proceeds via an excited state in $^{22}$O is unlikely
because the first excited state in $^{22}$O is located at an excitation energy of 3.2~MeV. It would place the narrow
resonance at an excitation energy of $~$6~MeV in $^{23}$O and it would be hard to reconcile the absence of any low energy unbound states in the spectrum.

\begin{figure}[tb!]
\begin{center}
\includegraphics[totalheight=6cm]{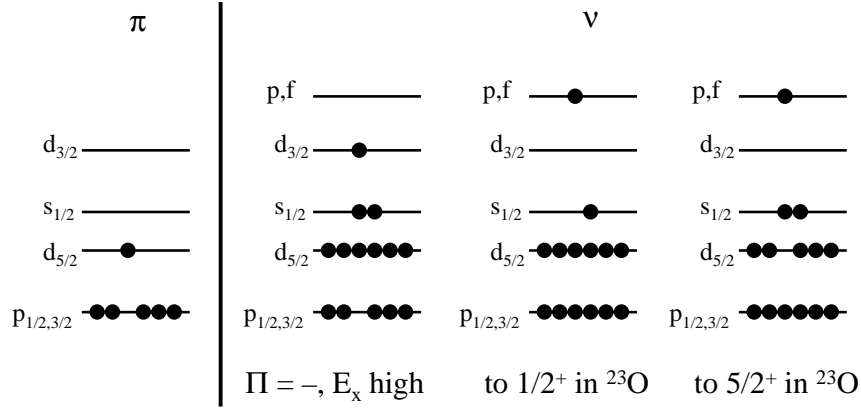}
\end{center}
\caption{Admixture of proton excitations and neutron excitations. The left side shows the knockout of the core
($p$-shell) proton. The neutron level diagrams on the right show three possible scenarios for cross shell neutron
excitations; see text for explanation.} \label{fig:admix}
\end{figure}

The observation of only one resonance in the decay-energy spectrum can be explained because the most straightforward reaction mechanism directly determines the spin assignment of the state. The ground-state wave function of the
secondary beam of $^{26}$Ne has large spectroscopic overlap with the $1/2^+$ ground state of $^{23}$O (a
$\nu(1s_{1/2})^{-1}$ hole state) \cite{Ter06} and the $5/2^+$ excited state, but the cross section for this
three-nucleon knockout reaction is rather small. A more likely scenario is the two-proton knockout involving one
valence $\pi(0d_{5/2})$ and one core $\pi(0p)$ proton. (The knockout of both valence protons ($\pi(0d_{5/2})^2$) will
simply populate the $^{24}$O ground state.) These excitations decay via negative-parity neutron-excitation admixtures
as shown in Figure \ref{fig:admix}. The proton configuration ($\pi(0p)+\pi(0d_{5/2})^{-1}$) shown on the left mixes
with (from left to right) neutron excitations of the form $\nu(0p)^{-1}\times\nu(0d_{3/2})^1$,
$\nu(1s_{1/2})^{-1}\times\nu(0f1p)^1$, and $\nu(0d_{5/2})^{-1}\times\nu(0f1p)^1$. The first of which has large
spectroscopic overlap with high-lying negative-parity excitations in $^{23}$O; the other two will excite one neutron
(either a $1s_{1/2}$ or a $0d_{5/2}$) to the continuum of the {\it fp}-shell. The current setup is not efficient for the detection of these high-energy neutrons which probably contribute to the
non-resonant background described above. The resulting $^{23}$O is then in the $1/2^+$ ground state and the $5/2^+$
excited state, respectively. The subsequent neutron decay of the $5/2^+$ state is the resonance observed in our
experiment.
Although it is not possible to distinguish between the three-nucleon knockout and the two-proton knockout followed by
neutron-emission, neither mechanism populates the $3/2^+$ (particle) excited state. It should be mentioned that this
state has recently been observed for the first time in the single-neutron transfer reaction $^{22}$O(d,p)$^{23}$O$^*$
\cite{Ele07}.

\section{Conclusions}\label{concl}
In conclusion, we have observed a 45(2)~keV resonance in the $n$-$^{22}$O decay-energy spectrum which we interpret as
the $5/2^+$ first excited state of $^{23}$O. The observation is consistent and complementary with the recent
observation of the $3/2^+$ state by Elekes {\it et al.} \cite{Ele07}. The non-observation of any higher-lying resonances implies selectivity of the
relevant reaction mechanism: either a direct three-nucleon removal or a core plus valence two-proton removal followed
by neutron emission.

\section*{Acknowledgments}
We would like to thank the members of the MoNA collaboration G. Christian, C. Hoffman, K.L. Jones, K.W. Kemper, P.
Pancella, G. Peaslee, W. Rogers, S. Tabor, and about 50 undergraduate students for their contributions to this work. We
would like to thank R.A. Kryger, C. Simenel, J.R. Terry, and K. Yoneda for their valuable help during the experiment.
Financial support from the National Science Foundation under grant numbers PHY-01-10253, PHY-03-54920, PHY-05-55366,
PHY-05-55445, and PHY-06-06007 is gratefully acknowledged. J.E.F. acknowledges support from the Research Excellence
Fund of Michigan.

\section*{Note}
\begin{notes}
\item[a]
Permanent address: Laboratoire de Physique Corpusculaire,
ENSICAEN, IN2P3, 14050 Caen, Cedex, France
\end{notes}

\vfill\eject
\end{document}